\documentclass[12pt]{article}
\begin{document}
\title{High Energy Dirac Solutions: Issues and Ramifications}
\author{B.G. Sidharth\\
International Institute for Applicable Mathematics \& Information Sciences\\
Hyderabad (India) \& Udine (Italy)\\
B.M. Birla Science Centre, Adarsh Nagar, Hyderabad - 500 063
(India)}
\date{}
\maketitle
\begin{abstract}
In this paper we consider solutions of the Dirac equation at ultra high energies. The study provides new insights including features
overlooked thus far and also new ramifications.
\end{abstract}
\section{Introduction}
It is known that at very high energies, we encounter negative
energies. This is because the set of positive energy solutions of
the Dirac or Klein-Gordon equations is not a complete set
\cite{feshbach}. At usual energies we could apply the well known
Foldy-Wothyson transformation to recover a description in terms of
positive energies alone or more precisely a description free of
operators which mix negative energy and positive energy components
of the wave function. This description also leads in the
non-relativistic limit to the two component Pauli equation
\cite{bd}. The situation is rather different at very high energies as we see below.
\section{Solutions at Ultra High Energies}
At very high energies the Cini-Toushek transformation of the Dirac equation leads to \cite{schweber}
\begin{equation}
H \psi = \frac{\vec{\alpha} \cdot \vec{p}}{|p|} E(p)\label{ex}
\end{equation}
where
\begin{equation}
\alpha^k = \left(\begin{array}{ll} 0 \quad \sigma^k\\
\sigma^k \quad 0\end{array}\right) \quad \quad \beta = \left(\begin{array}{ll} I \quad 0\\
0 \quad -I\end{array}\right)\label{24}
\end{equation}
\begin{equation}
\gamma^0 = \beta\label{26}
\end{equation}
More conventionally we have
\begin{equation}
\gamma^k = \beta \alpha^k \quad (k = 1,2,3)\label{27}
\end{equation}
where $\sigma^k$ are the Pauli matrices and $I$ is the $2 \times 2$
unit matrix.\\
We will also require the transformation of the $\gamma_5$ operator,
which is, given by,
\begin{equation}
\gamma_5 = \gamma^0 \gamma^1 \gamma^2 \gamma^3 = \imath \left(\begin{array}{ll} I \quad 0\\
0 \quad -I\end{array}\right)\label{86b}
\end{equation}
In the transformation (\ref{ex}) $\gamma_5$ goes over to $\Gamma_5$ given by
\begin{equation}
\Gamma_5 = \gamma_5 + \left(\frac{m}{E}\right) \left(\vec{\gamma}
\cdot \vec{n}\right) \gamma_5\label{III}
\end{equation}
In the above $\vec{n}$ is the unit vector in the direction of the
momentum vector. We can see from (\ref{III}) that
\begin{equation}
\Gamma_5 = \gamma_5\label{IV}
\end{equation}
whenever $m$ is negligible compared to $E:0 \left(\frac{m}{E}\right) < < 1$.\\
Returning to the Dirac equation we have:
\begin{equation}
\left(\gamma^\mu p_\mu - m\right) \psi = 0\label{8Hex2}
\end{equation}
It may be mentioned that two component \index{spinor}spinors
belonging to the representation \cite{schweber},
$$D^{\frac{1}{2}0} \, \mbox{or} \, D^{0\frac{1}{2}}$$
of the \index{Lorentz}Lorentz group are solutions of the
\index{Dirac}Dirac equation (\ref{8Hex2}). But these are no longer
invariant under reflections \cite{heine}. It is to preserve this
invariance that we have to consider the $4 \times 4$ representation
$$D^{(\frac{1}{2}0)} \oplus  D^{(0 \frac{1}{2})}$$
Under reflections, the two \index{spinor}spinors transform into each
other thus maintaining the overall invariance \cite{schweber}.\\
A similar analysis was carried out by Sudarshan \cite{sudarshan} and co-workers who considered separately
the usual Dirac regime denoted by $D$ and the high energy regime denoted by $E$. They defined projection
operators
\begin{equation}
\Lambda_{\pm}^D = \frac{1}{2} (1 \pm H^D/E), \quad (\Lambda_{\pm}^D)^2 = \Lambda_{\pm}^D,\label{5D}
\end{equation}
and
\begin{equation}
\Lambda_{\pm}^E = \frac{1}{2} (1\pm \alpha \cdot p/p).\label{5E}
\end{equation}
They then obtained the following relations for the coordinates $x,y,z$ marked by subscripts $+,-$ denoting the
positive energy and negative energy solutions:
$$[x_\pm , y_\pm ] = \left(\frac{\imath p_z}{2p^3} \gamma_5 \Lambda_\pm^E\right)_{E\,repres.}$$
\begin{equation}
\quad \quad \quad = \left(\pm \frac{\sigma_z}{2\imath p^2} \Lambda_\pm^D\right)_{D\,repres.}\label{su11}
\end{equation}
The remarkable result of (\ref{su11}) which was overlooked, was the fact that the coordinates no longer define a commutative spacetime.
\section{Noncommutative Spacetime}
Indeed Newton and Wigner \cite{newton,bgsextn} showed that the correct physical coordinate operator is given by
\begin{equation}
x^k = (1 + \gamma^0) \frac{p_0^{3/2}}{(p_0 + \mu)^{1/2}} \left(-\frac{\imath \partial}{\partial p_k}\right) \frac{p_0^{-1/2}}{(p_0 + \mu)^{1/2}} E\label{ext1} \end{equation}
where $E$ is a projection operator is given by
$$E = \frac{1}{2} p_0 (E\gamma^k p_k + \mu ) \gamma^0$$
and the gamma denote the usual Dirac matrices.\\
To appreciate the significance of (\ref{ext1}), we first consider the case of spin zero.\\
Then (\ref{ext1}) goes over to
\begin{equation}
x^k = \imath \frac{\partial}{\partial_{pk}} + \frac{1}{8\pi} \int \frac{exp(-\mu |(x-y|)}{|x-y|} \frac{\partial}{\partial y} dy\label{ext2}
\end{equation}
The first term on the right side of (\ref{ext2}) denotes the usual position operator, but the second term represents an imaginary part, which has an
$\sim 1/\mu$, the Compton wavelelength, exactly as in the case of the Dirac electron. Herein can be seen the origin of the
Compton scale.\\
Returning to Dirac's treatment of the electron \cite{diracpqm}, the position coordinate is given by
\begin{equation}
\vec{x} = \frac{c^2pt}{H} + \frac{1}{2} \imath c \hbar (\vec{\alpha} - c\vec{p} H^{-1}) H^{-1} \equiv \frac{c^2p}{H} t + \hat{x}\label{ext3}
\end{equation}
$H$ being the Hamiltonian operator and $\alpha$'s the non-commuting Dirac matrices, given by (\ref{24}).\\
The first term on the right hand side is the usual (Hermitian) position. The second term of $\vec{x}$ is the small oscillatory term of the order of
the Compton wavelength, arising out of zitterbewegung effects which averages out to zero. In other words to extension $l$ or the Compton
scale reappears.\\
On the other hand, if we were to work with the (non Hermitian) position operator in (\ref{ext3}), then we can easily verify that the following non
commutative geometry holds,
\begin{equation}
[x_\imath , x_j] = \alpha_{\imath j} l^2\label{ext4}
\end{equation}
where $\alpha_{\imath j} \sim 0$ (\ref{ex}). The $l^2$ in (\ref{ext4}) arises from the fact that the imaginary or non Hermitian part of
(\ref{ext3}) is an effect at the Compton scale $l$. In any case the non commutativity of the coordinates reappear.\\
While a relation like (\ref{ext4}) has been in use recently, in non commutative models including the author's own, and as will be noted below, was an independent starting point
due to the work of Snyder, we would like to stress that it has been overlooked that the origin of this non commutativity lies in the original Dirac coordinates (\ref{ext3}).\\
The relation (\ref{ext3}) shows that
\begin{equation}
c\vec{\alpha} = \frac{c^2\vec{p}}{H} - \frac{2\imath}{\hbar} \hat{x} H\label{ext5}
\end{equation}
In (\ref{ext5}), the first term is the usual momentum. The second term is the extra ``momentum" (and therefore energy) $\vec{p}$ due to the relations (\ref{ext4}).\\
In fact we can easily verify from (\ref{ext5}) that
\begin{equation}
\vec{\hat{p}} = \frac{H^2}{\hbar c^2} \hat{x}\label{ext6}
\end{equation}
where $\hat{x}$ has been defined in (\ref{ext3}).\\
Indeed in recent years this has lead to the so called Snyder-Sidharth dispersion relation \cite{tduniv}, wherein the energy takes on a small correction
term viz.,
$$E^2 = p^2 + m^2 + \alpha \frac{l^2}{\hbar^2} p^4$$
\section{Further Developments}
We now observe that \cite{bgsnap} with the transformation
\begin{equation}
\chi_j = E_j + \imath B_j, \chi_0 = 0\label{8Hex3}
\end{equation}
the Maxwell equations go over to \cite{barut}
\begin{equation}
\beta_\mu \frac{\partial \chi_\nu}{\partial x_\mu} = - \frac{1}{c}
j_\nu\label{8Hex4}
\end{equation}
which is in the form of the Dirac equation. In other words the zero mass or Cini-Toushek Dirac equation
described above gives back Maxwell's equations. We could say that the photon or the ultra high energy
electron is in a sense a combination of a neutrino and anti neutrino, both of which are described by two
component spinors except that we have to think in
terms of a combination given by
\begin{equation}
D^{(\frac{1}{2}0)} \oplus  D^{(0 \frac{1}{2})}\label{anb}
\end{equation}
rather than the regular bound state (Cf. Section 5).\\
Another way in which the above Dirac-Maxwell isomorphism can be realized is taking the dot product of the two divergence equations
of Maxwell with $\vec{\sigma}$ \cite{salhofer}. We then recover the above analysis as seen in Section 5.\\
The Feshbach and Villars interpretation throws further light \cite{feshbach}. Given
\begin{equation}
\Psi = \left(\begin{array}{ll} \phi \\
\chi\end{array}\right),\label{2.16}
\end{equation}
the Klein-Gordon equation can be written as
$$\imath \hbar (\partial \phi /\partial t) = (1/2m) (\hbar /\imath
\nabla - eA/c)^2 (\phi + \chi)$$
$$\quad \quad \quad +(e A_0 + mc^2)\phi,$$
\begin{equation}
\imath \hbar (\partial \chi / \partial t) = - (1/2m) (\hbar / \imath
\nabla - eA/c)^2 (\phi + \chi) + (e A_0 - mc^2)\chi\label{2.15}
\end{equation}
\cite{bgsultra}.
We note that (\ref{2.15}) represent two coupled equations for the two component objects $\phi$ and $\chi$. If however, $e = 0 = m$ in (\ref{2.15}),
these get uncoupled, giving the Klein-Gordon (and Dirac) equations for $\phi$ and $\chi$ separately, representing ``uncoupled"
neutrinos and antineutrinos. On the other hand, a non-zero mass implies mixing of positive and negative solutions and vice-versa.\\
More generally this leads to the following scenario: For one observer we have,
\begin{equation}
\Psi \sim \left(\begin{array}{ll} \phi \\
0\end{array}\right)\label{B}
\end{equation}
and for another higher energy observer we have
\begin{equation}
\Psi \sim \left(\begin{array}{ll} 0 \\
\chi\end{array}\right)\label{C}
\end{equation}
This is what may be called the mathematical bound state given in (\ref{anb}).\\
The above scenario would be valid for a single observer if a particle velocity got a
sudden large boost. All this is true of any Dirac particle. This shows that there is a
transmutation between particles and antiparticles as indeed has been observed, for example for the
$B$-meson (Cf.ref.\cite{bgs}).\\
We can see the above from another point of view. Our starting point is the solution of the
Dirac equation \cite{bd} given by
$$\psi(x,t) = \int \frac{d^3p}{(2\pi \hbar)^{1/2}} \sqrt{\frac{mc^2}{E}} \sum_{\pm s} \left[b(p,s)u(p,s)e^{-\imath p^{\mu} x_{u}/\hbar}\right.$$
\begin{equation}
\quad \quad \quad \left.+ d^* (p,s)v(p,s)e^{+\imath p^{\mu} x_{\mu}/\hbar} \right].\label{1}
\end{equation}
where the current vector is given by
$$J^k = \int d^3p \left\{ \sum_{\pm s} [|b(p,s)|^2 + |d(p,s)|^2] \frac{p^kc^2}{E}\right.$$
$$\quad \imath \sum_{\pm s, \pm s'} \, b^*(-p,s')d^*(p,s)e^{2\imath x_0 p_0/\hbar_{\bar{u}}}(-p,s')\sigma^{k0}v(p,s)$$
\begin{equation}
\left.-\imath \sum_{\pm s, \pm s'} \, b(-p,s')d(p,s)e^{-2\imath x_0 p_0/\hbar_{\bar{v}}}(p,s')\sigma^{k0}u(-p,s)\right\}.\label{2}
\end{equation}
As can be seen from (\ref{2}), there is the time independent group velocity given by the first term within paranehesis.
Additionally there is a sum of two cross products of positive and negative energy solutions. This cross product as
can be seen oscillates very rapidly with the frequency $\frac{2 p_0c}{\hbar}$. It is this rapid oscillation that is the Zitterbewegung
which needs to be eliminated. For a localized wave packet there are two regions: an inner region of the order of the Compton
scale within which the packet is packed and on the region outside. It is in this region, which we may call the outer region that our usual concepts of spacetime appear. Within this region
however as can be seen from (\ref{2}) we have a mixture of eigen states of positive energy and negative energy that is we have a
situation where time oscillates between positive and negative values rapidly. Indeed Wigner and Salecker showed that there can be
no physical or definite time within the Compton scale and that time as we know it is a phenomenon outside this scale.\\
The Compton scale gives rise to the Zero Point Field which in the author's work models dark energy \cite{cu}. The positive and negative eigen
values which arise from (\ref{2}) for the $\frac{d}{dt}$ or energy operator inside the Compton wavelength clearly define a non differentiable
time rapidly oscillating (Cf. also ref.\cite{ichi}).\\
Another way of looking at this is, following Sudarshan and co-workers, we write the wave function as
$$\psi = u_+ + u_-$$
where the right side represents positive and negative energy solutions. As we saw,
Sudarshan then got \cite{sudarshan} a noncommutative geometry, which unfortunately was overlooked.\\
The above clearly shows that the energy or $d/dt$ operator does not have a fixed eigen value but rather due to the
Zitterbewegung rapidly oscillates with eigen value $+E$ and $-E$ represents in the Feshbach-Villars representation a rapid flip flop
between particles and antiparticles, or as all this happens in the Compton Scale, it could be interpreted as in recent times, as
a particle being surrounded by a cloud of extremely short lived particles and antiparticles. This as we know leads to renormalization techniques.\\
We could think of the above in terms of the Wiener process. In this case the probability density
is given by
\begin{equation}
fW_t (x) = \frac{1}{\sqrt{2\pi t^2}} e^{-\frac{x^2}{t^2}}\label{ea}
\end{equation}
whose expectation is zero:
\begin{equation}
E[W_t] = 0.\label{ec}
\end{equation}
The variance however is given by
\begin{equation}
Var(W_t) = E[W^2_t] - E^2[W_t] = E[W^2_t] - 0 = E[W^2_t] = t^2\label{eb}
\end{equation}
It turns out as can be seen from (\ref{eb}) that the time $t$ in this formulation appears as a standard deviation. This was argued several
years ago by the author, directly from a Random Walk process \cite{bgsheap}. Thus time appears with a slightly different character. With
$N$ events which can be likened to $N$ random tosses or steps, the fundamental interval between events being given by $\tau$, we have the time elapsed
$T$ given by
$$T = \sqrt{N} \tau$$
One could see that this Wiener process in the Zitterbewegung region, that is the Compton scale clearly leads to a break of symmetry as argued
for the Kaon and the $B$-meson \cite{bgscurrentscience}. This could be characterized as a new interaction defined by the covariant
derivative which by Ito's lemma can be written as
\begin{equation}
df (X_t) = f' (X_t) dX_t + \frac{1}{2} f^{''} (X_t) \sigma^2_t dt.\label{ed}
\end{equation}
where the subscript $t$ refers to the instant of time in question. (\ref{ed}) though stochastic, resembles an electromagnetic (gauge) interaction
as $\sigma$, the standard deviation is a scalar. On the other hand this arises from the Zero Point Field. This interaction in the model can be looked upon as causing the
flip between particles and antiparticles and vice versa.\\ \\
We also note that the zero point energy encountered above comes from the normal modes of oscillators and is $\approx \omega/2$. As is known \cite{bd}\\
such a fluctuating electromagnetic field would persist even when there are no external fields. Even though the average strength is zero, there
is the residual mean square value for this field and that is the Zero Point Field whose effect can be seen for instance in the well
known Lambshift of energy levels.\\
We can even show that classical physics with the addition of Zero Point Field leads to the Quantum Mechanical commutation relations, that
is Quantum Theory itself \cite{sachi,tduniv}.\\
It is also known that it is the Zero Point Field which leads to other Quantum effects like the anomalous $g = 2$ factor. Furthermore, as can be seen from
(\ref{ext2}) and (\ref{ext3}), this field also gives rise to noncommutative spacetime.\\
We may note that it is fluctuations that underpin the universe. In the author's 1997 model \cite{uof,mg8} it was this Zero Point Field,
later called Dark Energy that caused the slightly accelerated expansion of the universe which was subsequently confirmed in 1998 by the
observations of Perlmutter, Schmidt and Riess \cite{perl}.\\
We also note that once we introduce the non differentiable time as above, then rather than the Schrodinger formulation that is usually obtained, we
can directly get the Klein-Gordon relativistic case. This has been described in detail \cite{tduniv}. But briefly the origin of this is a
treatment of discrete states like the two state system which now leads to
\begin{equation}
C_\imath (t - \Delta t) - C_\imath (t + \Delta t) = \sum_{j}
\left[\delta_{\imath j} - \frac{\imath}{\hbar} H_{\imath
j}(t)\right] C_j^{(t)} \Delta t\label{2ge}
\end{equation}
whence the Klein-Gordon equation follows.\\ \\
{\bf Matter, Antimatter, Asymmetry}: This has been one of the unsolved puzzles because it is widely believed that at the time of
the Big Bang particles and antiparticles were created in equal numbers. Yet matter dominates in the universe overwhelmingly. The above
considerations throw light on this. This can be seen from equations (\ref{B}) and (\ref{C}). In the initial epoch the negative energy
solution or antiparticle (\ref{C}) would dominate but with the rapid cooling of the temperature (\ref{B}) would begin to dominate.
As noted this would imply a flip flop from antiparticles to particles. It has already been pointed out by the author \cite{ijmpe} that
this process could be designated in terms of a new super spin: Thus one state would be a super spin up and the other would be a super spin down. This
has been observed for $B$-mesons as noted above.\\
The other aspect is that as argued by the author \cite{bgsejtp}, there is an inherent slight asymmetry between particle solutions and
antiparticle solutions. For a slightly different treatment (Cf.ref.\cite{bgsneutrino}). Indeed in recent years there have been indications, particularly from the
MINOS experiment of Fermi Lab that indeed neutrinos and antineutrinos not only have slightly different behaviour, but also masses.
\section{Brief Remarks}
It will be evident from the above considerations that the Dirac sea has been replaced by the positive and negative energy solutions or swarm of
particles and antiparticles within the Compton wavelength. We have also seen that with a sudden and steep increase or decrease of energy equations
(\ref{B}) and (\ref{C}) show the possibility of transmutation between antiparticles and particles. This could explain as noted why our universe as it cooled down
transited into one of the particles rather than antiparticles.\\
In fact in the above analysis the Dirac wave functions
are given by, apart from multiplicative factors
\begin{equation}
\left[\begin{array}{ll} 1 \\ 0 \\ \frac{p_zc}{E + mc^2} \\
\frac{p+c}{E + mc^2}
\end{array}\right]
\left[\begin{array}{ll} 0 \\ 1 \\ \frac{p-c}{E + mc^2} \\
\frac{-p_zc}{E + mc^2}
\end{array}\right]
\left[\begin{array}{ll} \frac{p_z c}{E + mc^2} \\ \frac{p + c}{E +
mc^2} \\ 1 \\ 0
\end{array}\right]
\left[\begin{array}{ll} \frac{p+c}{E+mc^2} \\ \frac{-p_z c}{E +
mc^2} \\ 0 \\ 1
\end{array}\right]\label{8Hex5}
\end{equation}
At very high energies (\ref{8Hex5}) are simplified into
\begin{equation}
\psi_1 = \left[\begin{array}{ll} 1 \\ 0 \\ 1\\ 0
\end{array}\right]
\psi_2 = \left[\begin{array}{ll} 0 \\ 1 \\ 0 \\ -1
\end{array}\right]
\psi_3 = \left[\begin{array}{ll} 1 \\ 0 \\ 1 \\ 0
\end{array}\right]
\psi_4 = \left[\begin{array}{ll} 0 \\ -1 \\ 0 \\ 1
\end{array}\right]
\label{8Hex6}
\end{equation}
These do not constitute a couple of neutrino equations. This is because of the relation which we have assumed namely
\begin{equation}
\chi_j = E_j + \imath B_j, \chi_0 = 0\label{8Hex3}
\end{equation}
which effectively couples the two solutions. (But Cf. the discussion after equation (\ref{2.15})).
These must now be treated as the up and down states of the super spin, rather than a bound state.
\newpage
\section{Final Remark}
We have already referred to the Maxwell-Dirac isomorphism. Let us start with the charge free Maxwell equations
$$div E = 0,$$
\begin{equation}
div B = 0.\label{sal13}
\end{equation}
We then get, on taking the scalar product with the Pauli matrices (Cf.ref.\cite{salhofer})
\begin{equation}
\left\{\begin{array}{ll}
(\sigma \cdot \nabla)(\sigma \cdot.H) - \frac{\epsilon}{c} \frac{\partial}{\partial t} (\imath \sigma \cdot E) = 0.\\
(\sigma \cdot \nabla)(\sigma \cdot E) + \frac{\mu}{c} \frac{\partial}{\partial t} (\sigma \cdot H) = 0.
\end{array}\right\}\label{sal15}
\end{equation}
In matrix notation this reads
\begin{equation}
\left[\left(\begin{array}{ll}
1 \quad \sigma \\
\sigma \quad 0
\end{array}\right) \cdot \nabla - \left(\begin{array}{ll}
\epsilon 1 \quad 0 \\
0 \quad \mu 1
\end{array}\right)
\frac{1}{c} \frac{\partial}{\partial t}\right] \left[
\begin{array}{ll}
\imath (\sigma \cdot E) \\
(\sigma \cdot H)
\end{array}\right] = 0.\label{sal16}
\end{equation}
We get for (\ref{sal16}) using
\begin{equation}
\Psi = \psi e^{-\imath \omega t}\label{sal20}
\end{equation}
\begin{equation}
\left[ \gamma \cdot \nabla + \imath \frac{\omega}{c} \left(\begin{array}{ll}
\epsilon 1 \quad 0\\
0 \quad \mu 1
\end{array}\right)\right] \Psi = 0.\label{sal21}
\end{equation}
This is in agreement with the Dirac equation
\begin{equation}
\left[\gamma \cdot \nabla + \imath \frac{\omega}{c} \left(\begin{array}{ll}
\left(1 - \frac{\Phi - m_0c^2}{\hbar \omega}\right)1 \quad \quad \quad 0\\
0 \quad \quad \left(1 - \frac{\Phi + m_0c^2}{\hbar \omega}\right)1
\end{array}\right)\right] \Psi = 0\label{sal22}
\end{equation}
That is the multiplication of the charge free Maxwell equations by the Pauli spin vector leads to the
Dirac equation.


\begin{thebibliography}{99}
\bibitem {feshbach} Freshbach, H. and Villars, F. (1958).
\emph{Rev.Mod.Phys.} Vol.30, No.1, January 1958, pp.24-45.
\bibitem {bd} Bjorken, J.D. and Drell, S.D. (1964). \emph{Relativistic Quantum Mechanics}
(Mc-Graw Hill, New York), pp.39.
\bibitem {schweber} S.S. Schweber. (1961). \emph{An Introduction to Relativistic Quantum
Field Theory} (Harper and Row, New York).
\bibitem {heine} Heine, V. (1960). \emph{Group Theory in Quantum Mechanics} (Pergamon Press,
Oxford), pp.364.
\bibitem {sudarshan} Sudarshan, E.C.G., Bose, S.K. and Gambu, A. (1959). \emph{Phys.Rev.} 113 (6) March 1959.
\bibitem {newton} Newton, T.D. and Wigner, E.P. (1949). \emph{Reviews of Modern Physics} Vol.21,
No.3, July 1949, pp.400-405 --26-1.
\bibitem {bgsextn} Sidharth, B.G. (2002). \emph{Foundation of Physics
Letters}, 15(5),2002,501ff.
\bibitem {diracpqm} Dirac, P.A.M. (1958). \emph{The Principles of
Quantum Mechanics} (Clarendon Press, Oxford), pp.4ff, pp.253ff.
\bibitem {tduniv} Sidharth, B.G. (2008). \emph{Thermodynamic Universe} (World
Scientific, Singapore, 2008).
\bibitem {bgsnap} Sidharth, B.G. (2013). \emph{New Advances in Physics} Vol 7 (2), 2013.
\bibitem {barut} Barut, A.O. (1964). \emph{Electrodynamics and Classical Theory of Fields and
Particles} (Dover Publications, New York), p.97ff.
\bibitem {salhofer} Salhofer, H.H. in \emph{Essays on the Formal Aspects of Electromagnetic Theory} Ed.Akhlesh Lakhtakia
(World Scientific, Singapore, 1993), pp.268ff.
\bibitem {bgsultra} Sidharth, B.G. (2005). \emph{Int.J.Mod.Phys.E.}
14(6),2005,927ff.
\bibitem {bgs} Sidharth, B.G. \emph{arXiv 1107.4459}.
\bibitem {cu} Sidharth, B.G. (2001). \emph{Chaotic Universe: From the Planck to the Hubble Scale}
(Nova Science, New York).
\bibitem {ichi} Ichinose, T. (1984). \emph{Physica} 124A, pp.419.
\bibitem {bgsheap} Sidharth, B.G. (2000). \emph{Chaos, Solitons and Fractals}
12(1), 2000, 173-178.
\bibitem {bgscurrentscience} Sidharth, B.G. (2013). \emph{Current Science} March 25, 2013.
\bibitem {sachi} Sachidanandam, S. (1983). \emph{Physics Letters} Vol.97A, No.8,
19 September 1983, pp.323--324.
\bibitem {uof} Sidharth, B.G. (1998). \emph{Int.J. of Mod.Phys.A} 13, (15), pp.2599ff.
\bibitem {mg8} Sidharth, B.G. (1999). \emph{Proc. of the Eighth Marcell Grossmann Meeting on
General Relativity (1997)} Piran, T. (ed.) (World Scientific,
Singapore), pp.476--479.
\bibitem {perl} Perlmutter, S.,  et al. (1998). \emph{Nature} Vol.391, 1 January 1998, pp.51--59.
\bibitem {ijmpe} Sidharth, B.G. (2011). \emph{Int.J.Mod.Phys.E} 20 (10), 2011, pp.2177-2188.
\bibitem {bgsejtp} Sidharth, B.G. (2010). \emph{El.J.Th.Phys.} Vol.7, No.24, July 2010.
\bibitem {bgsneutrino} Sidharth, B.G. and Glinka, L.A. (2013). \emph{On Neutrino and Anti-Neutrino Solutions
in Modified Relativity} in \emph{GPBC-01-01-2013}.
\end{thebibliography}
\end{document}